\documentclass[doublecol]{epl2} 
\usepackage{amsmath,amssymb} \usepackage{colordvi} \usepackage{color}
\usepackage{graphicx}
\usepackage{dcolumn}
\usepackage{bm}
\usepackage{romannum}
\usepackage{caption}
\usepackage{subcaption}
\usepackage{cleveref}
\usepackage{float}
\usepackage{csquotes}

\title{Studies of a single component Fermi gas near a $p$-wave resonance with the lowest order constrained variational method }
\shorttitle{Studies of the single component Fermi gas near a $p$-wave resonance} 

\author{Guangcun Liu\inst{1} \and Yi-Cai Zhang\inst{2} }
\shortauthor{Guangcun Liu and Yi-Cai Zhang}

\institute{
  \inst{1} Department of Physics and Center of Theoretical and Computational Physics, The University  of Hong Kong\\
   Hong Kong, China\\
  \inst{2} Department of Physics, College of Physics and Electronic Engineering, Guangzhou University\\
   Guangzhou 510006, China
}

\abstract{
 We study a single component Fermi gas near a $p$-wave resonance with the lowest order constrained variational (LOCV) method. We obtain the energy per particle for the ground state of single component Fermi gas near a $p$-wave resonance with LOCV method. We also calculate compressibility of the single component Fermi gas near a $p$-wave resonance and it shows that in the strongly interacting BCS side, the system would lose its stability and collapse. The width of unstable region is proportional to Fermi energy which reflects the universality of dilute Fermi gas. The two $p$-wave contacts are also obtained  and their variation tendencies with the interaction strength are qualitatively in agreement with recent experimental results.}

\begin{document}

\maketitle

\section{\label{sec:level1}Introduction}

The $p$-wave interaction between particles has attracted considerable attention for a long time in condensed matter physics.
 The interest in the $p$-wave system mainly originates from the distinguished features of the superfluid phase of Helium-3~\cite{Leggett,Vollhardt,Chubukov,helium3}.
 Recently the single component Fermi gas near a $p$-wave resonance has become a topic of intense research both experimentally \cite{Regal2003,Zhang2004,Ticknor,Gaebler2007,Inada2008,Fuschs2008,Nakasuji2013,Waseem2016,Onofrio} and theoretically. For example, the possible superfluid phases in $p$-wave channels have been investigated \cite{Ho2005,Botelho,Ohashi,Gurarie,melo,Levinsen2007}.  The anisotropic superfluid phase may appear due to the dipolar interaction between atoms \cite{Cheng}. Comparing with usually broad s-wave Feshbach resonance interaction, the $p$-wave Feshbach resonance is usually narrow where the closed channel molecules have large weights in scattering wave function \cite{Gubbels}. For $p$-wave scattering, the correction of van der Waals' forces to the effective range expansion in phase shift has been discussed \cite{zhangpeng}. Due to its relatively large weight of two-body wave function near origin, the molecule-molecule collision can form trimmer or deeply bound molecules. It is shown that
the lifetime of a three-dimensional $p$-wave superfluid is much shorter than its $s$-wave counterpart \cite{Jona-Lasinio,Levinsen2008}.

In ultracold atomic physics, another important research topic is the universal properties near $s$-wave Feshbach resonance \cite{tan1,platter, tan2,tan3,scontact,castin,unitaryfermigas}.  Recently, the above ideas on the universal properties have been generalized to high partial wave cases. For example, the concepts of so called $p$-wave contacts have been proposed \cite{ueda,pcontact,zhou,hu} and experimentally observed \cite{Luciuk}, which can be used to characterize some universal properties of strongly interacting $p$-wave gas.  The physics near the Feshbach resonance has become an interesting and challenging topic because of the lack of traditional perturbation parameter and hence can't be treated with traditional perturbation theory.

The lowest order constrained variational (LOCV) method~\cite{locv1,locv2,locv3} can be used in calculating the ground state energy of a strongly interacting quantum fluid. It has already been used in calculating the Bose gas with large scattering length \cite{bose}. The energy per particle for the ground state of two-component Fermi gas near a $s$-wave resonance has also been calculated with the LOCV method~\cite{monte,2006phd} and the results obtained are in agreement with quantum Monte Carlo calculation. The LOCV method has the advantage that it is much simpler in numerical calculation. The energy per particle for Bose and two-component Fermi system with higher partial waves is also obtained with the LOCV method although $p$-wave calculation is unphysical because the trial wave function does not obey the correct statistical properties of identical particles~\cite{ryan}.

Although the thermodynamic potential for single component $p$-wave Fermi gas at finite temperature (near the superfluidity transition point) \cite{juan,Inotani} has been calculated, the explicit calculations on the ground state energy for such system are rare. Motivated by the recent theoretical and experimental progresses in $p$-wave resonance~\cite{Luciuk,ueda,pcontact,zhou,hu},
we adopt the LOCV method to examine the single component Fermi gas system near $p$-wave resonance and discuss its stability. It would be convenient to extract the compressibility of the system near $p$-wave resonance since the ground state energy can be calculated accurately with the LOCV method. Another essential quantity, the $p$-wave contact~\cite{ueda,pcontact,zhou,hu}, can also be obtained with the LOCV method.

\section{\label{sec:level1}Model and method}
We apply the lowest order constrained variational (LOCV) method \cite{locv1,locv2,locv3} to a single component $p$-wave Fermi gas.
The Hamiltonian of the system is:
\begin{equation}
\hat{H}=-\frac{\hbar^2}{2M}\sum_{i}^{N}\nabla_{i}^2+\sum_{i<j}^{N}v_{i,j},
\end{equation}
where $M$ is particle mass and $v_{i,j}$ is two-body interaction potential, $N$ is total particle number. In the following, we set $\hbar=M\equiv1$.
The many-body Schr\"{o}dinger equation is
\begin{equation}
\left[-\frac{1}{2}\sum_{i}^{N}\nabla_{i}^2+\sum_{i<j}^{N}v_{i,j}\right]\Psi(1\cdots N)=E\Psi(1\cdots N), \label{manyse}
\end{equation}
where $E$ is the total energy of the many-body system.

In order to solve the above eq. (\ref{manyse}), we choose the following trial wave function for a single component fermion,
\begin{equation}
\begin{aligned}
&\Psi(1\cdots N)= \\
&\frac{1}{\sqrt{N!}}\sum_{P}\delta_{p}P\prod_{i<j}f_{ij}(|\vec{r}_i-\vec{r}_j|,|\vec{k}_i-\vec{k}_j|)\varphi_1\cdots\varphi_N.
\label{wavefunction}
\end{aligned}
\end{equation}
$\sum_{P}$ denotes the summation over all the permutations ($P$) of  $N$ particle's position $\vec{r}_1,\vec{r}_2,...,\vec{r}_N$. $\delta_{p}=1(-1)$ for even(odd) permutation and $\varphi_i=e^{i\vec{k}_i\cdot\vec{r}_i}$ is the single particle plane wave function. $f_{ij}(|\vec{r}_i-\vec{r}_j|,|\vec{k}_i-\vec{k}_j|)$ is the Jastrow factor \cite{jastrow}. If $f_{ij}=1$, the above wave function eq.~(\ref{wavefunction}) is just a Slater determinant which describes the non-interacting Fermi gas.
So $f_{ij}(|\vec{r}_i-\vec{r}_j|,|\vec{k}_i-\vec{k}_j|)$ describes the modification of non-interacting fermion wave function due to the two-body interaction.
For single component Fermi gas any pair of fermions is in the spin triplet state. We set $f_{ij}$ to be a function of the absolute value of relative position and relative momentum, so that the trial wave function would obey the correct quantum statistic symmetry when two fermions are exchanged. As the variational parameters, $f_{ij}$ need to be determined by minimizing the energy of the system.

The expectation value of $\hat{H}$ (ground state energy) can be calculated with the trial wave function $\Psi$,
\begin{equation}
E=\frac{\left\langle \Psi| \hat{H} |\Psi \right\rangle }{\left\langle\Psi|\Psi\right\rangle}.
\end{equation}
$\Psi$ includes all possible correlations. The LOCV method tries to simplify the calculation by only considering the two-body correlations and neglects three-order or higher-order correlations, i.e. \cite{locv3},
\begin{equation}
\text{if}\ f_{ij}\neq1, \text{then}\  f_{ik}=1 \ \text{for}\ k \neq j.
\label{boundarycondition1}
\end{equation}

What is more, when one considers the short-range interaction, the correlations between two particles are weak if they are far apart, so $f_{ij}$ tends to unity (non-interacting value). Thus one can impose a constrain on $f_{ij}$~\cite{locv3}:
\begin{equation}
f_{ij}(r>d)=1,\  f'_{ij}(r=d)=0,
\label{boundarycondition2}
\end{equation}
$d$ is called the healing length which needs to be determined self-consistently. $d$ is usually of order of the inverse of Fermi momentum $k_F$.
Within the distance $d$, the  function $f_{ij}$ deviates from non-interacting value, i.e., $f_{ij}(r<d)\neq1$.

With the trial wave function eq.~(\ref{wavefunction}) and the two conditions eq.~(\ref{boundarycondition1}) and eq.~(\ref{boundarycondition2}), one could obtain a concise expression for the energy of the system. Defining wave functions for relative motion $\Psi_{ij}(r)=f_{ij}\psi_{ij}$ and $\psi_{ij}=e^{i(\vec{k}_{i}-\vec{k}_{j})\cdot\frac{\vec{r}}{2}}$ (we drop the subindex $(i,j)$ of $\Psi_{i,j}$ hereafter), the expression for the energy per particle can be written as:
\begin{equation}
\begin{aligned}
\frac{E}{N}-\frac{E_0}{N}
=\frac{1}{2NV}\sum_{k_{i}<k_{j}}\int_0^d[\Psi(\vec{r})^{*}-\Psi(-\vec{r})^{*}]\times \\
(v-k^2- \nabla^2)\times[\Psi(\vec{r})-\Psi(-\vec{r})]d^3\vec{r}
,
\end{aligned}
\label{3denergy}
\end{equation}
where $\sum_{k_{i}<k_{j}}$ denotes the summation over all the possible pairs of two momenta below the Fermi sphere. $E_0$ is the kinetic energy of non-interacting Fermi gas, $k^2=[\frac{1}{2}(\vec{k}_i-\vec{k}_j)]^2$ and $k$ is relative momentum of a pair of particles. Then, our task will be to minimize the total energy under the constrain of eq.~(\ref{boundarycondition2}).

It is convenient to decompose $\Psi$ in partial waves
\begin{equation}
\Psi(\vec{r})=\sum_{l.m}R_l(kr)Y_{l,m} (\theta,\phi),
\end{equation}
where $Y_{l,m}$ is spherical harmonic function  and $R_l(kr)=f^{k}_{l}(r) j_l(kr)$  is radial wave function of a pair of particles, $j_{l}(kr)$ is spherical Bessel function describing the non-interacting gas while $f^{k}_{l}(r)$ gives corrections due to interactions. For identical fermions, the angular momentum quantum number $l$ must be odd integer.
From eq.~(\ref{3denergy}), the variational equation for $R_l(kr)$ becomes:
\begin{equation}
-\frac{d^2 R_l}{dr^2}-\frac{2}{r}\frac{dR_l}{dr}+\frac{l(l+1)}{r^2}R_l+vR_l-k^2R_l=\lambda_0^l(k)R_l.
\label{eigenequation}
\end{equation}
Here $\lambda_0^l(k)$ is the energy correction due to two-body correlation.
We see that if $v=0$ (non-interacting limit), the wave function reduces to non-interacting Bessel function and the energy correction $\lambda_0^l(k)=0$.

\begin{figure}[ht]
\centering
\includegraphics[scale=0.40]{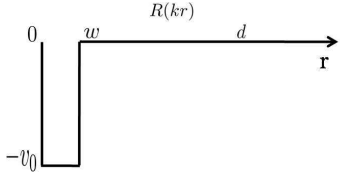}
\caption{Square well potential with well depth $v_0$ and well width $w$. $d$ is the healing length which is much larger than $w$. $R(kr)$ is the radial wave function and needs to follow the boundary conditions at $r=w$ and $r=d$.}
\label{square}
\end{figure}

Further more, we take a model potential, i.e., square well potential with well depth $v_0$ and well width $w$ as shown in Fig. \ref{square}. In the dilute limit, the healing length would be much larger than potential width, i.e., $d\gg w$.
We set $w$ as the unit for length and $1/w$ as the unit for wave vector hereafter.

In the following, we don't consider the effect of interaction on higher partial waves which means $f^{k}_{l}$ is set to be 1 for higher partial waves ($l>1$). This is reasonable near $p$-wave resonance because the effect of interaction on other odd partial waves is much smaller than effect on $p$-wave.
Then, the wave function for $r<w$ can be written as: $R(kr)\equiv R_{1}(kr)=Aj_1(k_1r)$ with $k_1^2=\lambda_0^1+v_0+k^2$. $j_1(k_1r)$ is spherical Bessel function. For $w<r<d$, the wave function is $R(kr)=Bj_1(k_2r)+Cn_1(k_2r)$ with $k_2^2=\lambda_0^1+k^2$. $n_1(k_2r)$ is the spherical Neumann function. For $r>d$, $R(kr)=i\sqrt{12\pi }j_1(kr)$ is the result of the constrain eq.~(\ref{boundarycondition2}). Then, there are four boundary conditions since the radial wave function $R(kr)$ and its first derivative need to be continuous at $r=w$ and $r=d$.

The healing length $d$ is determined by the normalization condition \cite{locv1,locv2,locv3}:
\begin{equation}
\frac{1}{NV}\sum_{k_i,k_j}\int_0^d[\Psi(\vec r)^{*}-\Psi(-\vec r)^{*}][\Psi(\vec r)]d^3r=1,\label{norm}
\end{equation}
which can be simplified as :
\begin{equation}
\begin{aligned}
&\frac{4}{\pi^4n}\int_0^{k_F}dkk^2\left[\frac{k_F^3}{3}+\frac{k^3}{6}-\frac{kk_F^2}{2}\right]\int_0^d[R(kr)]^2r^2dr  \\
&=1,
\end{aligned}\label{normalization}
\end{equation}
where $V$  is volume of system, $k_F$ is Fermi momentum and  the density $n=k_F^3/6\pi^2$.

From the above normalization condition, we could see the difference between $p$-wave case and $s$-wave case for the LOCV method \cite{locv3,ryan}. For any paired particles in $p$-wave case, the relative momentum is non-zero while the relative momentum is zero for $s$-wave case. Then the normalization for $p$-wave case needs to average all the possible relative momenta between paired particles and the weight for relative momentum $k$ in the normalization relies on integration of the radial wave function $[R(kr)]^2$ from the space $\vec{r}=0$ to $\vec{r}=d$. This difference is also reflected in the expression of interaction energy eq.~(\ref{interaction}) (please see below).

To be accurate, we also consider higher odd partial waves' contribution in the radial wave integration part of the normalization condition.  Then one can add higher odd partial waves' contribution on the integration of radial wave function on the left hand side of  the normalization equation (eq.~(\ref{normalization})), which is:
\begin{equation}
\begin{aligned}
&\sum_{l\neq1,l \text{ is odd}}\int_{0}^{d}4\pi(2l+1)j_l^2(kr)r^2dr= \\
&2\pi\int_0^d \left[1-\frac{\cos(kr)\sin(kr)}{kr}\right]r^2dr-12\pi\int_0^d  j_1^2(kr)r^2dr.
\end{aligned}
\end{equation}
Then, the boundary condition eq.~(\ref{boundarycondition2}), normalization eq.~(\ref{normalization}) and the eigenequation eq.~(\ref{eigenequation}) can be solved iteratively to determine the healing length $d$ and then to calculate the energy of the system through eq.~(\ref{3denergy}).

The energy correction from non-interacting Fermi kinetic energy per particle is:
\begin{equation}
\begin{aligned}
&E_{I}/N= \\
&\frac{2}{\pi^4n}\int_0^{k_F}dkk^2\left[\frac{k_F^3}{3}+\frac{k^3}{6}-\frac{kk_F^2}{2}\right]\lambda_0^1(k)\int_0^d[R(kr)]^2r^2dr.
\end{aligned}\label{interaction}
\end{equation}

The non-interacting Fermi kinetic energy per particle is:
\begin{equation}
E_0 /N=1/N\sum_kk^2/2=3/5E_F
\end{equation}
with $E_F=k_F^2/2$.

%

Until now, the lowest order constrained variational method for calculating the energy per particle near $p$-wave resonance has been presented. Once the energy of a system is obtained, the related thermodynamic quantities can also be calculated.
\section{\label{sec:level1}$p$-wave scattering of two atoms}
In order to clarify the problem, we review some background knowledge about two-body $p$-wave scattering problem.
In three dimensions, the expansion of $\cot\delta_1$ ($\delta_1$ is the $p$-wave phase shift) for small relative momentum $k$ can be written as~\cite{landau}:
\begin{equation}
k^3\cot\delta_1=-\frac{1}{v}-\frac{k^2}{R}+\mathcal{O}(k^4),
\end{equation}
$v$ is the scattering volume and $R$ the effective range. They are
functions of scattering potential depth and width. Note here we refer the \enquote{resonance}
as $1/v=0$ and thus the resonance position does not depend on relative momentum. But this is different from the usual definition of resonance scattering which requires the real part of denominator of scattering amplitude equals zero ($k^2v+R=0$) and hence should be dependent on the relative momentum in $p$-wave case. In the following, we will see that the relative momentum dependence of the resonance in $p$-wave case turns out to be essential in the understanding of our results obtained with the LOCV method which we show later in this paper.

\begin{figure}[ht]
\centering
\includegraphics[scale=0.42]{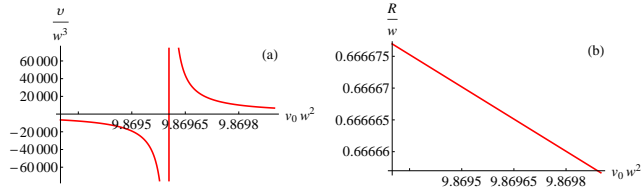}
\caption{(a) Scattering volume $v$ and (b) effective range $R$ as functions of attractive trap depth near $p$-wave resonance.}
\label{RV}
\end{figure}

We show the behavior of $v$ and $R$ near $p$-wave resonance in Fig.~\ref{RV}. Here we just consider the most shallow attractive trap depth that could form a two-body bound state. We are only interested in the strongly interacting regime which can be characterized with the dimensionless parameter $|vk_{F}^2/R|\sim1$ (analogously to $|k_Fa_s|\sim1$ for $s$-wave case with $a_s$ scattering length). As shown in Fig.~\ref{RV}, when the attractive trap depth is smaller than the resonant potential depth $v_0=9.8696/w^2$ (BCS side), the scattering volume is negative and approaches negative infinity when the attractive trap depth goes near to the resonant potential depth. Once the potential depth is larger than the resonant potential depth (BEC side), the scattering volume becomes positive infinite and then decreases as the potential depth increases. In contrast, the effective range is smooth and positive definite through the resonance.

The scattering amplitude can be written as:
\begin{equation}
f_1=\frac{1}{g-ik},
\end{equation}
with $g=k\cot\delta_1$
and $k^3\cot\delta_1=-1/v-k^2/R$, then
\begin{equation}
f_1=\frac{1}{-\frac{1}{k^2v}-\frac{1}{R}-ik}.
\end{equation}

We introduce
\begin{equation}
E_b\equiv R/v,
\end{equation}
and $-E_b=-R/v=0.0$ corresponds to the $p$-wave resonance. For $-E_b=-R/v>0$, the system is in the BCS side and when $-E_b=-R/v<0$ the system is in the BEC side. So we can take $-E_b$ as the tunable energy scale to characterize the $p$-wave resonance.

In addition, from condition of true resonance scattering ($k^2v+R=0$), one can easily see that the unitary limit (maximum of scattering cross section) should appear at BCS side where $v<0$. One would expect that an unstable region for single-component Fermi gas, if any, should occur at strongly interacting BCS side, i.e.,  $v<0$ and $|vk_{F}^2/R|\sim1$.  As we will see in the following, this is indeed the case.

\section{\label{sec:level1}Energy}
\begin{figure}[ht]
\centering
\includegraphics[scale=0.60]{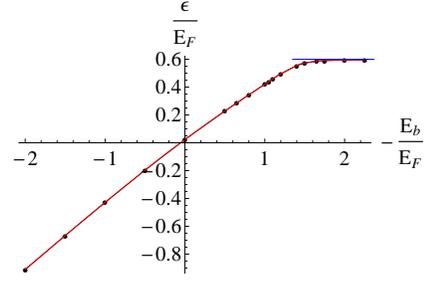}
\caption{Energy per particle near $p$-wave resonance for three dimensions.  Fermi energy $E_F=k_F^2/2$ with $k_F=0.01/w$. The black dots are calculated data, the red line is to guide the eye. The blue line is the average energy per particle for non-interacting single component fermions. }
\label{energy}
\end{figure}

With the LOCV method, we calculate the energy per particle near $p$-wave resonance. The result is shown in Fig.~\ref{energy}.
As shown in Fig.~\ref{energy}, in BCS side, when $-E_b=-R/v$ is large (approaching the right side of figure), the energy per particle is very close to the average energy of a fermion in the non-interacting single component fermion system ($0.6E_F$). Attractive interaction could reduce the energy but the modification is very small. Near resonance, the attractive interaction starts to obviously reduce the energy and the energy decreases smoothly through the resonance to the BEC side. In the deep BEC side ($-E_b/E_F\ll -1$), the energy per particle is very close to the energy of half of the two-body bound state energy (out of the figure on the left side). Similar results are also obtained in earlier work with different methods~\cite{Ho2005,Pricoupenko}.

\section{\label{sec:level1}Compressibility}
Two-component Fermi gases have been confirmed to be stable \cite{sstable} and their universal properties have been examined extensively near s-wave resonance~\cite{sreview}. The stability of bosons and boson-fermion mixture near Feshbach resonance also have been examined carefully~\cite{boson,mixture}.  With the ground state energy obtained from the LOCV calculation, it is straightforward to calculate the compressibility $\kappa$ for single component $p$-wave case.

Compressibility $\kappa$ is the relative volume change against a pressure change for the fixed particle number $N$,
\begin{equation}
\begin{aligned}
\kappa&=-\frac{1}{V}\left( \frac{\partial V}{\partial P}\right)\bigg\vert_{N}=V^{-1}\left(\frac{\partial^2E}{\partial V^2}\right)^{-1}\bigg\vert_{N}.
\end{aligned}
\end{equation}

In the thermodynamic limit, $E=(n\epsilon)V$ where $\epsilon$ is the energy per particle and $n$ is the particle density. Hence
\begin{equation}
\begin{aligned}
\frac{1}{\kappa}&=n^2\left(\frac{d^2(\epsilon n)}{dn^2}\right)=n^2\left(2\frac{\partial \epsilon}{\partial n} +n\frac{\partial^2 \epsilon}{\partial n^2} \right).
\end{aligned}
\end{equation}

For a non-interacting Fermi gas at zero temperature, the chemical potential is
\begin{equation}
\begin{aligned}
\mu_0=\frac{k_F^2}{2}
\end{aligned}
\end{equation}
and inverse of compressibility is
\begin{equation}
\frac{1}{\kappa_0}=n^2\left(\frac{d\mu_0}{dn}\right)=\frac{2\pi^2n^2}{k_F}.
\end{equation}
\begin{figure}[ht]
\centering
\includegraphics[scale=0.7]{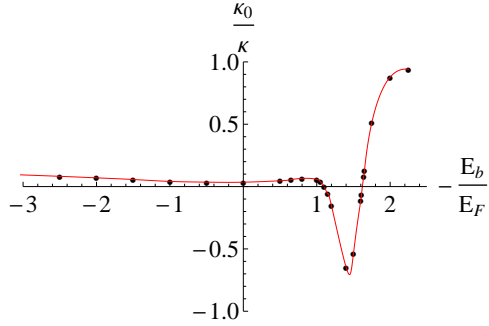}
\caption{Compressibility $\kappa$ near $p$-wave resonance. $\kappa_0$ is the compressibility for non interacting single component Fermi gas. Fermi energy $E_F=k_F^2/2$ with $k_F=0.01/w$. The black dots are calculated data, the red line is to guide the eye.}
\label{kappa}
\end{figure}

We report  inverse compressibility $1/\kappa$ normalized by $1/\kappa_0$ in Fig.~\ref{kappa}.  When $-E_{b}/E_{F}$ is large, $\kappa$ is very close to the compressibility of non-interacting Fermi gas as expected. But around $-E_{b}/E_{F}=1.625$, $\kappa_0/\kappa$ approaches zero and then becomes negative until $-E_{b}/E_{F}=1.1$. When $-E_{b}/E_{F}$ is smaller than 1.1, $\kappa_0/\kappa$ becomes positive again. In the region $1.1<-E_b/E_{F}<1.625$, compressibility is negative which indicates that the system is not stable. It is worth mentioning that this region corresponds to the region where the interaction starts to obviously modify the energy per particle from the average energy of non-interacting Fermi gas and the second order derivative of energy per particle over density changes dramatically (see Fig. \ref{energy}). Crossing over this region, the compressibility becomes positive again and the system becomes stable. We note the instability occurs in the strongly interacting BCS region where $|k_{F}^2v/R|=-2E_F/E_b\approx 1.23-1.82$.

\begin{figure}[!h]
\centering
\includegraphics[scale=0.4]{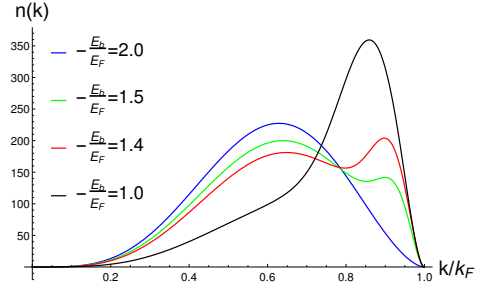}
\caption{Normalized weight of relative momentum ($k_F=0.01/w$). }
\label{nk}
\end{figure}

For the negative compressibility obtained with the LOCV method, we provide more discussions in order to clarify it. We calculate the normalized weight of relative momentum for different interaction strength near the unstable region. The definition for the normalized weight of relative momentum is:
\begin{equation}
n(k)=\frac{4}{\pi^4n}k^2\left[\frac{k_F^3}{3}+\frac{k^3}{6}-\frac{kk_F^2}{2}\right]\int_0^d[R(kr)]^2r^2dr.
\end{equation}
Compared with the eq.~(\ref{normalization}), the physical meaning is the weight of relative momentum in normalization. The results for some typical cases are shown in Fig. \ref{nk}.

It is clear that the weight for large relative momentum increases when the system approaches the unstable region. As discussed earlier, if we define `resonance' as that the real part of scattering amplitude is zero, the resonance position becomes relative momentum dependent and larger relative momentum reaches its resonance earlier than smaller relative momentum when the interaction strength increases from a weakly interacting regime to a strongly interacting regime. As for the specific cases shown in Fig. \ref{nk}, the resonance momentum for $-E_b/E_F=2.0$ is $k=k_F$, which lies on the right edge of the weight of relative momentum with vanishingly small weight; for $-E_b/E_F=1.5$, the resonance momentum is $k=0.866025k_F$ which has very large distribution weight; for $-E_b/E_F=1.4$, the resonance momentum is $k=0.83666k_F$, which also has very large distribution weight, but for $-E_b/E_F=1.0$, the resonance momentum is $k=0.7071k_F$, which lies on the left side of its maximum of weight of relative momentum and hence has smaller distribution weight. We also find similar behavior for other smaller values of $-E_b/E_F$, i.e., the normalized weight of relative momentum has large weight on the large relative momentum while the resonance momentum becomes smaller and further away from the maximum value of the weight and thus has smaller weight. From the above description, we could observe that the momentum dependent resonance are in fact important especially when the relative momentum has different weight in the normalized weight of relative momentum. For the unstable region as shown in Fig. \ref{kappa}, the resonant relative momentum is between $0.9k_F$ and $0.75k_F$ which also lies very close to the maximum of corresponding normalized weight of relative momentum. Thus the large weights for resonance momenta result in the instability.

Meanwhile, this argument can also be used to explain the stability when $-E_b/E_F$ is very close to zero where two-body scattering volume goes to infinity. This is because the corresponding resonance momentum is very close to zero. Because the resonance momentum has vanishingly small weight in the normalized weight of relative momentum, the compressibility should be positive and the system is stable.

\begin{figure}[ht]
\centering
\includegraphics[scale=0.6]{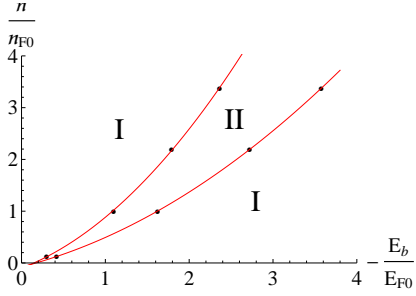}
\caption{Region I is the stable region and region II is the unstable region for different densities $n$.  $E_{F0}=k_{F0}^2/2$ and $n_{F0}=k_{F0}^3/6\pi^2$ with $k_{F0}=0.01/w$. The black dots are calculated data, the red line is to guide the eye.}
\label{densitydependent}
\end{figure}

We also examine the density dependence of unstable region. The result is shown in Fig.~\ref{densitydependent}.
As shown in Fig.~\ref{densitydependent}, when density decreases, the unstable region approaches to the $p$-wave resonance point ($v\rightarrow\infty$). The simple reason for this is that when the particle density decreases, $k_F $ decreases, in order to reach strongly interacting region ($vk^2_{F}/R\sim1$), the corresponding $v$ needs to become larger ($R$ can be regarded as constant here).  At the same time, we find that our result shows that the width for unstable region is $\Delta E_b\simeq C E_F$ where $C$ is density-independent constant which is approximately 0.525. This behavior is a manifestation of universality for the $p$-wave dilute Fermi gas. With dimensional analysis, the energy per particle can be written as $\epsilon=E_{F}f(v,R,k_F)=E_{F}g(k_FR, R/vk_{F}^{2})$, where $f$ and $g$ are some dimensionless functions. Due to the smallness of $k_FR$  for dilute Fermi gas ($k_FR\ll1$), the energy $\epsilon$ in the unstable regime should be a single-variable
function of $R/vk_{F}^{2}$ or $-E_b/E_F$.
i.e., $\epsilon=E_{F}g(-E_b/E_F)$.
Consequently the width of unstable region
over Fermi energy, i.e., $\Delta E_b/E_F$ should not depend on density any more and thus we have $\Delta E_b\propto E_F$.

In a recent paper \cite{zhoufei}, stability of two-dimensional $p$-wave superfluidity near $p$-wave resonance is examined and the instability of $p$-wave superfluid near $p$-wave resonance due to quantum fluctuations is also revealed.

\section{\label{sec:level1}$p$-wave contact}
The $s$-wave contact has been a central quantity in considering a two-component dilute Fermi gas system with s-wave scattering \cite{ho,platter,tan1,tan2,tan3,scontact,castin}. The theories for $p$-wave contact have already been proposed~\cite{ueda,pcontact,zhou,hu} and have received considerable interest~\cite{cui1,cui2,juan,twodcontact}. Unlike s-wave contact, there are two $p$-wave contacts which are related to the scattering volume and effective range in three dimensions. Both contacts have already been measured in a recent experiment with fully polarized Fermi gas $^{40}K$~\cite{Luciuk}. With the LOCV method, we calculate the two $p$-wave contacts for single component fermions.

The two $p$-wave contacts are defined as:
\begin{equation}
\frac{dE}{dv^{-1}}\bigg|_R=-\frac{1}{2}\sum_{m}C_{v}^{(m)}, \ \ \frac{dE}{dR^{-1}}\bigg|_v=-\frac{1}{2}\sum_{m}C_{R}^{(m)},
\end{equation}
$v$ is the scattering volume, $R$ the effective range. $m$ is the projection of angular momentum onto the quantized axis. In our case, they are degenerate since there is no external magnetic field.

The two three-dimensional $p$-wave contacts can be calculated as following:

\begin{equation}
\begin{bmatrix}
\sum_{m}C_{v}^{(m)} \\
\sum_{m}C_{R}^{(m)}
\end{bmatrix}
=-2\left[ \begin{array}{cc}
\frac{\partial v^{-1}}{\partial v_{0}} & \frac{\partial R^{-1}}{\partial v_{0}} \\
\frac{\partial v^{-1}}{\partial w} & \frac{\partial R^{-1}}{\partial w} \end{array} \right]^{-1}\begin{bmatrix}
\frac{\partial (n\epsilon)}{\partial v_{0}} \\
\frac{\partial (n\epsilon)}{\partial w}
\end{bmatrix}.
\end{equation}

Since we can calculate the energy per particle quite accurately, we can then obtain the two contacts. We carefully checked the accuracy of single particle energy and it is reliable for the calculation of contacts.  \newline
\begin{figure}[H]
\centering
\includegraphics[scale=0.45]{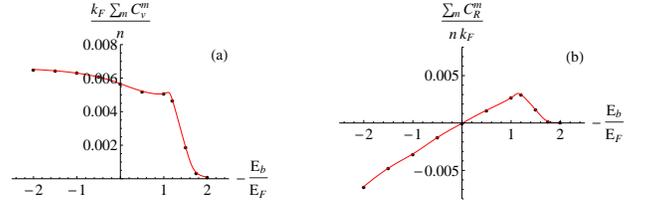}
\caption{(a) $\sum_{m}C_{v}^{(m)}$ near $p$-wave resonance. (b) $\sum_{m}C_{R}^{(m)}$ near $p$-wave resonance. For both figures Fermi energy $E_F=k_F^2/2$ with $k_F=0.01/w$. The black dots are calculated data, the red line is to guide the eye.}%
\label{crandcv}
\end{figure}

The results of the two contacts are shown in Fig. \ref{crandcv}. $C_v$ increases monotonically while $C_R$ increases first and then decreases as $-E_b/E_F$ decreases. The maximum value for $C_R$ appears at $-E_b/E_F$ around 1.2. It is interesting that the region where $C_v$ and $C_R$ increases significantly corresponds to the unstable region in Fig.~\ref{kappa}. In addition, our results are qualitatively in agreement with the experimental measurements \cite{Luciuk}.


\section{\label{sec:level1}Summary}
To summarize, we extend the lowest order constrained variational (LOCV) method to $p$-wave case and use this method to study the single component Fermi gas near a $p$-wave resonance. The ground state energy near a three-dimensional $p$-wave resonance is calculated and the calculated compressibility near a $p$-wave resonance shows that there is a region where the system would lose its stability.
 The variation tendencies of $p$-wave contacts in three dimensions are qualitatively in agreement with experimental results. We emphasize that, since the energy for dilute Fermi gas has universal properties, the quantity and variation trends derived from energy, e.g., compressibility and two $p$-wave contacts could also have universal properties. At last, we should remark that we only consider the lowest order terms and ignore higher order terms since the calculation of higher order terms is much more complicated, it's possible that higher order terms could modify our results. It's necessary to use more exact method such as quantum Monte Carlo method to obtain more precise results of this system.

\acknowledgments
We thank Shizhong Zhang and Zhenhua Yu for useful discussions.
This work is supported by Hong Kong Research
Grants Council (General Research Fund, HKU 17318316 and Colaborative Research Fund, C6026-16W), and NSFC under Grants
No.11747079. The startup grant from Guangzhou University (Y.C.Zhang) is gratefully acknowledged.

\end{document}